\documentclass[twocolumn,showpacs,preprintnumbers,amsmath,amssymb]{revtex4}
\reversemarginpar
\usepackage{labelfig,graphicx}

\usepackage{graphicx}% Include figure files
\usepackage{dcolumn}% Align table columns on decimal point
\usepackage{bm}% bold math
\usepackage{amsfonts}
\usepackage{latexsym}

\begin{document}

\preprint{APS/123-QED}

\newcommand{\commentout}[1]{}
\newcommand{\nwc}{\newcommand}
\nwc{\rer}{r_{\eta,\rho}}
\nwc{\corr}{r_{\eta,\rho}}
\nwc{\rinf}{r_{\eta,\infty}}
\newcommand{\myphi}{\Phi_{(\eta,\rho)}}
\nwc{\xvec}{{\vec{\bx}}}
\nwc{\kvec}{{\vec{\bk}}}
\newcommand{\lt}{\left}
\newcommand{\rt}{\right}
\newcommand{\vas}{\varepsilon}
\newcommand{\lan}{\left\langle}
\newcommand{\ran}{\right\rangle}
\newcommand{\tvas}{\Psi^\vas}
\newcommand{\pdgx}{\bp\cdot\nabla_\bx}
\newcommand{\psiep}{\Psi^\vas}
\newcommand{\vvas}{\tilde{\ml V}_z^\vas}
\newcommand{\veptil}{\tilde{\ml V}_z^\vas}
\newcommand{\vep}{{\ml V}_z^\vas}
\newcommand{\cv}{{\ml V}^\ep_z}
\newcommand{\cvtil}{\tilde{{\ml V}}^\ep_z}
\newcommand{\n}{\nabla}
\newcommand{\tkappa}{\tilde\kappa}
\newcommand{\ks}{k}
\newcommand{\cs}{\tilde{c}_0}
\newcommand{\bx}{\mathbf x}
\newcommand{\bp}{\mathbf p}
\newcommand{\by}{\mathbf y}
\newcommand{\bv}{\mathbf v}
\newcommand{\bq}{\mathbf q}
\newcommand{\bw}{\mathbf w}
\newcommand{\br}{\mathbf r}
\newcommand{\bs}{\mathbf s}
\newcommand{\bH}{\mathbf H}
\newcommand{\bE}{\mathbf E}

\newcommand{\one}{1\hspace{-4.4pt}1}

% theorem-like enviroments:

\nwc{\nwt}{\newtheorem}
\nwt{assumption}{Assumption}
%\nwt{prop}{Proposition}
%\nwt{proposition}{Proposition}
%\nwt{lemma}{Lemma}
%\nwt{theorem}{Theorem}
\nwt{cor}{Corollary}
%\nwt{remark}{Remark}
%\nwt{definition}{Definition} %def is already defined

%\nwc{\ba}{\begin{array}}
\nwc{\bal}{\begin{align}}
\nwc{\be}{\begin{equation}}
%\nwc{\be}{\begin{eqnarray}}
\nwc{\ben}{\begin{equation*}}
\nwc{\bea}{\begin{eqnarray}}
\nwc{\beq}{\begin{eqnarray}}
\nwc{\bean}{\begin{eqnarray*}}
\nwc{\beqn}{\begin{eqnarray*}}
\nwc{\beqast}{\begin{eqnarray*}}

%\nwc{\ea}{\end{array}}
\nwc{\eal}{\end{align}}
\nwc{\ee}{\end{equation}}
%\nwc{\ee}{\end{eqnarray}}
\nwc{\een}{\end{equation*}}
\nwc{\eea}{\end{eqnarray}}
\nwc{\eeq}{\end{eqnarray}}
\nwc{\eean}{\end{eqnarray*}}
\nwc{\eeqn}{\end{eqnarray*}}
\nwc{\eeqast}{\end{eqnarray*}}

\nwc{\tx}{\tilde{\bx}}
\nwc{\tp}{\tilde{\bp}}
\nwc{\tr}{\tilde{\br}}
\nwc{\tw}{\tilde{\bw}}
\nwc{\ep}{\varepsilon}
\nwc{\ept}{\epsilon}
\nwc{\vrho}{\varrho}
\nwc{\orho}{\bar\varrho}
\nwc{\ou}{\bar u}
\nwc{\vpsi}{\varpsi}
\nwc{\lamb}{\lambda}
\nwc{\wep}{W^\ep}

\nwc{\nn}{\nonumber}
%\nwc{\bm}{\boldmath}
\nwc{\mf}{\mathbf}
\nwc{\mb}{\mathbf}
\nwc{\ml}{\mathcal}

\nwc{\IA}{\mathbb{A}} %algebraic
\nwc{\IB}{\mathbb{B}}
\nwc{\IC}{\mathbb{C}} %complex
\nwc{\ID}{\mathbb{D}} %Dedekind
\nwc{\IM}{\mathbb{M}} %Dedekind
\nwc{\IP}{\mathbb{P}} %Dedekind
\nwc{\II}{\mathbb{I}} %Dedekind
\nwc{\IE}{\mathbb{E}} %Euklides
\nwc{\IF}{\mathbb{F}} %finite field
\nwc{\IG}{\mathbb{G}} %Gauss
\nwc{\IN}{\mathbb{N}} %natural
\nwc{\IQ}{\mathbb{Q}} %rational
\nwc{\IR}{\mathbb{R}} %real
\nwc{\IT}{\mathbb{T}} %torus
\nwc{\IZ}{\mathbb{Z}} %integers

\nwc{\cE}{{\ml E}}
\nwc{\cP}{{\ml P}}
\nwc{\cL}{{\ml L}}
\nwc{\cR}{{\ml R}}
\nwc{\cV}{{\ml V}}
\nwc{\cW}{{\ml W}}
\nwc{\cT}{{\ml T}}
\nwc{\crV}{{\ml V}_{(\delta,\rho)}}
\nwc{\cC}{{\ml C}}
\nwc{\cA}{{\ml A}}
\nwc{\cS}{{\ml S}}
\nwc{\cK}{{\ml K}}
\nwc{\cB}{{\ml B}}
\nwc{\cD}{{\ml D}}
\nwc{\cF}{{\ml F}}
\nwc{\cM}{{\ml M}}
\nwc{\cN}{{\ml N}}
\nwc{\cG}{{\ml G}}
\nwc{\cH}{{\ml H}}
\nwc{\bk}{{\mb k}}
\nwc{\cQ}{{\ml Q}}
\nwc{\cO}{{\ml O}}
\nwc{\cJ}{{\ml J}}

\nwc{\mint}{{\int\cdot\int}}

\title{Superresolution and 
Duality for  Time-Reversal of  Waves in Self-Similar Media}

\author{Albert Fannjiang}
 \email{cafannjian@ucdavis.edu.
}
 \affiliation{
Department of Mathematics,
University of California at Davis,
Davis, CA 95616 
} 
\author{Knut Solna}%
\email{ksolna@math.uci.edu.
}
\affiliation{
Department of Mathematics,
University of California at Irvine, Irvine CA 92697.}

\begin{abstract}
We analyze the time reversal of waves in a turbulent medium
using the parabolic Markovian model.  We prove that
the time reversal resolution can be a nonlinear function
of the wavelength and independent of the aperture. 
We establish a duality relation between the turbulence-induced wave spread and the time-reversal resolution which
can be viewed as an uncertainty inequality for random media.
The  inequality becomes an equality when
the wave structure function is Gaussian.
\end{abstract}

% PACS, the Physics and Astronomy

\maketitle

\section*{Introduction}
Time reversal is the process of recording the signal from
a remote source, time-reversing and back-propagating it
to retrofocus around the source. Time reversal of acoustic waves
 has been demonstrated to hold exciting
technological potentials  in subwavelength focusing, dispersion compensation, communications, imaging, remote-sensing and target detection in unknown environments (
see \cite{Fink} and references therein). 
The same should hold for the
 electromagnetic waves as well.
 Time reversal of electromagnetic waves is closely
 related to optical phase conjugation (OPC)
 which used to be limited to monochromatic
 waves. With the advent of
 experimental techniques, time reversal
 of high frequency EM waves hold 
diverse potential applications
 including real-time adaptive optics, laser resonators, high-power laser systems, optical communication and information processing, image transmission, spatial and temporal filtering, spectroscopy etc \cite{Fink}. 
 
 \commentout{
 Some of these promises are still unfulfilled. For example,  in spite of the amazing imaging properties of OPC conventional lens optics is still the dominating technique for fine-line lithography applications. This is mainly due to the many unsolved resolution problems of the nonlinear phase conjugator
 which is so far limited to processing a monochromatic wave.
 But the new experimental techniques with linear optical devices
 such as photonic crystals.
 }

Time reversal refocusing is the result of the time-reversal invariance
of 
the wave equations, acoustic or
electromagnetic, in  time invariant media.
The surprising and important
fact is that the refocal spot in a
richly scattering medium is typically  {\em smaller}
than that in the homogeneous medium.
That is, the time reversal resolution is enhanced 
rather than hampered by the inhomogeneities
of the medium. This sub-diffraction-limit retrofocusing is sometimes called
{\em superresolution} and in certain regimes has been explained
mathematically by using radiative transfer equations.

In the previous experimental, numerical or theoretical results
the superresolution comes as a {\em linear} function of
the wavelength but {\em independent} of the aperture.
In this letter
we show that in fractal media the resolution can
be a {\em superlinear } (between linear and quadratic) function of the wavelength and at the same time independent
of the aperture. The lowest achievable refocal spot size
in this nonlinear regime is on the order of the smallest scale of the medium
fluctuations. Below the inner scale  the resolution is diffraction-limited while 
above the outer scale it is the previously reported  aperture-independent enhanced resolution  \cite{Fink, BPZ}.

We will focus our analysis on the widely used {\em parabolic
Markovian} model for  waves in atmospheric turbulence
\cite{St}. Neglecting the depolarization effect let us
write the forward propagating wave field $E$ at the carrier
wave number $k$ as $E(z,\bx)=\Psi(z,\bx) e^{i(kz-\omega t)},
\bx\in \IR^2$
where the complex wave
amplitude $\Psi$ satisfies the 
 Schr\"ondinger equation in
 the non-dimensionalized form
\beq
i2\ks\frac{\partial \Psi}{\partial z} +
{\gamma} \Delta_\perp\Psi+\frac{\ks^2}{\gamma}
V\lt({z}, {\bx}\rt)\circ\Psi=0
\label{para}
\label{para2}
\eeq
with $\Delta_\perp$ being the Laplacian in
the transverse coordinates $\bx\in \IR^2$ and
$V$ the fluctuation of the refractive index.
Here  the Fresnel number $\gamma$ equals $L_z k_0^{-1}L_x^{-2} $ with $k_0$ being the reference
wavenumber, $L_z$ and $L_x$ the reference
scales in the longitudinal and transverse directions,
respectively. 
The notation $\circ$ in eq. (\ref{para}) means
the Stratonovich product (v.s. It\^o product). 
In the Markovian model $V(z,\cdot)$ is assumed to
be a $\delta$-correlated-in-$z$ stationary random field
such that 
\[
\lan V(z,\bx) V(z', \bx')\ran
=\delta(z-z')\int \Phi(0, \bp) e^{i\bp\cdot(\bx-\bx')}d\bp
\]
where $\Phi(\kvec), \kvec=(\xi,\bp)\in \IR^3$ is the power spectrum density of the refractive index fluctuation and,
in the case of atmospheric turbulence, has a
power-law behavior in the inertial range. For
simplicity of presentation we assume an 
isotropic 
power-law 
\beq
\label{power}
 \Phi(\kvec)=\sigma_H
|\kvec|^{-1-2H}|\kvec|^{-2},\quad |\kvec|\in
(L_0^{-1}, \ell_0^{-1})
\eeq
where $L_0$ and $\ell_0$ are respectively the
outer and inner scales of the turbulence  and 
 $\sigma_H$ a constant factor.
 Usually  $H$ is taken to be $1/3$ in
 the self-similar theory of turbulence.
 We assume that the spectrum decays sufficiently fast for
$|\kvec|\gg \ell_0^{-1}$ while staying bounded for $|\kvec| \ll
L_0^{-1}$.

For this model of propagation we will prove
an uncertainty inequality where the
conjugate quantities are the forward
wave spread and the time-reversal resolution.
The inequality becomes an {\em equality} when
the wave structure function is Gaussian.
This  
also helps illustrating  an experimentally observed, 
close relation 
between the time reversal resolution and
the correlation length of the scattered wave field
prior to time reversal \cite{Fink}.
%Hence the time-reversal resolution can be used
%to  provide
%a robust estimation of 
%the correlation length of the scatted wave field.
\section*{Time reversal process}
\begin{figure}
\begin{center}
\includegraphics[width=2.5in, totalheight=1.5in]{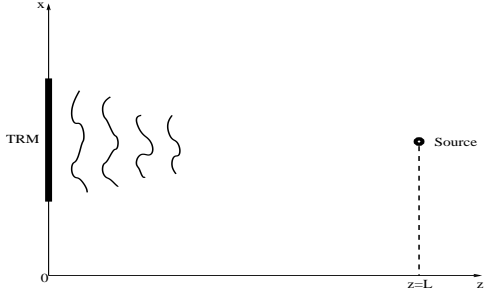}
\end{center}
\label{figback}
\caption{The time reversal process
 }
\end{figure}
In the time reversal procedure,  a source $\Psi_0(\bx)$ located at $z=L$ emits
a signal with 
the carrier wavenumber $k$ toward the time reversal
mirror (TRM) of aperture $A$  located at $z=0$ through a turbulent medium.
The transmitted field is captured and time reversed
at the TRM 
and then sent back toward the source point through
the same turbulent medium,  see Figure 1.

The time-reversed, back-propagated wave field  at $z=L$ can be
expressed as 
\begin{eqnarray} 
\label{eq:gpro} 
\lefteqn{\Psi_{\rm tr}(\bx)}\\
&=&
\int  
G(L, \bx, \bx_m)  
\overline{ G(L, \bx_s, \bx_m)
  \Psi_0\lt({\bx_s }\rt) }\II_A(\bx_m) d\bx_m
d\bx_s   \nn\\
&=&\int  
 e^{i \bp \cdot (\bx - \bx_s)/\gamma}
 W(L, \frac{\bx+\bx_s}{2},\bp)
   \overline{\Psi_0\lt({\bx_s }\rt)}d\bp d\bx_s
\label{eq:gpro2} 
\end{eqnarray}
where $\II_A$ is the indicator function of
the TRM, $G$ the propagator of
the Schr\"odinger equation and $W$ the Wigner distribution function
\bea
\label{ww}
{W(z,\bx,\bp)}
 &=& \nn \frac{1}{(2\pi)^d}\int e^{-i\bp\cdot\by} G(z,
\bx+\gamma\by/2,
\bx_m)\\
&&\times 
\overline{ G(z,\bx-\gamma\by/2, \bx_m ) }
\II_A(\bx_m)d\by d\bx_m.\nn
\eea
Here we have used the fact
that time reversing of the signal is equivalent to
the phase conjugating of its spatial component.

The Wigner distribution $W$ satisfies a closed form
equation, the Wigner-Moyal equation \cite{rad-arma},
and for the Markovian model its moments also satisfy
 closed form equations. In particular, the mean field
 equation is
 \beq
 \frac{\partial \lan W\ran }{\partial z}
 +\frac{\bp}{k}\cdot\nabla_\bx \lan W\ran =\cQ \lan W\ran
 \label{moyal}
 \eeq
with the scattering operator $\cQ$  given by
 \beq
\nn
 \cQ f(\bx,\bp)
 &=&\frac{k^2}{4\gamma^2}
 \int \Phi(0,\bq)\lt[-2 f(\bx,\bp)+f(\bp+\gamma \bq)\rt.\\
 &&\lt.
 +f(\bx,\bp-\gamma\bq)\rt]d\bq.
 \label{scatter}
 \eeq
 
 Eq. (\ref{moyal}) is exactly solvable
 and 
the mean refocused field  of the parabolic Markovian model
can be expressed as 
\bea
\label{eq:psi} 
\lefteqn{\lan \Psi_{\rm tr}\ran (\bx) }\\
 &=&\frac{1}{(2\pi)^{2}}
  \int d\bx' d\bq d\bw
   \overline{\hat{\Psi}_0 \left( {\bq}\right)}
      \II_A(\bx')\exp{\lt[i\bq\cdot\bx\rt]}\nn\\
      &&\times
\exp{\lt[i(\bw\cdot(\bx-\bx')
   -\gamma L\bw\cdot\bq/\ks-\gamma L|\bw|^2/2\ks)\rt]}
	  \nn  \\&&\times
\exp{\lt[-\ks^2/(2\gamma^2) \int_0^L {D_*}(-s\gamma\bw/\ks) ds\rt]}
\nn
\eea
where  the structure function ${D_*}$ is given by
\bea
   {D_*}(\bx) &=&  \int \Phi( 0, \bq)
        \lt[1-  e^{i\bx\cdot\bq}\rt]\, d\bq.
        \label{9}
\eea
Here and below $\hat f$ denotes $\cF f$ the Fourier transform
of $f$.
The main property of ${D_*}$ we need in the subsequent analysis
is the inertial range asymptotic:
\beq
\label{holder}
  {D_*}(r) \approx C_{*}^2 r^{2H_*},\quad
\ell_0\ll r\ll L_0 \, ,
\eeq
where the effective H\"older exponent $H_*$ is given by
\beq
\label{holder2}
  H_*=\left\{
\begin{array}{ll}
 H+1/2&\hbox{for}\,\, H\in (0,1/2)\\
1&\hbox{for}\,\,H\in (1/2,1]
\end{array}
\right.
\eeq
and the structure parameter $C_{*}$ is proportional
to $\sigma_H$.
Outside of the inertial range we have instead 
$
D_*(r)\sim r^2, r\ll \ell_0$ and $D_*(r)\to D_*(\infty)$ for $r\to\infty$
where $D_*(\infty)>0$ is a constant.

Let us consider a point source located at $(L,\bx_0)$ by substituting
the Dirac-delta function $\delta(\bx-\bx_0)$ for $\Psi_0$ in (\ref{eq:psi}).  
We then obtain  the  point-spread function for time reversal
$
\cP_{tr} =\cP_{0} T_{tr}
$
with
\bea\label{88}\nn
 \lefteqn{ {\cP}_0(\bx-\bx_0)}\\
  &\equiv&
\left(\frac{\ks}{\gamma L}\right)^2 e^{i\frac{k}{2\gamma L}(|\bx|^2-|\bx_0|^2)}\hat{\II}_A\left(\frac{\ks}{\gamma L}(\bx-\bx_0)\right)\nn \\
\lefteqn{T_{tr}(\bx-\bx_0)}\label{ttr}\\
&\equiv&  \exp{\lt[-\ks^2/(2\gamma^2) L\int_0^1
{D_*}(-s(\bx-\bx_0)) ds\rt]}.\nn
 \eeq
In the absence of random inhomogeneity 
the function $T_{\rm tr}$ is unity and the resolution
scale $\rho_0$
 is determined solely by ${\cP}_0$:
\beq
\label{ray}
\rho_0\sim \gamma\frac{{\lamb}L}{A},\quad
 {\lamb}=\frac{2\pi}{\ks} .
\eeq
This is the classical (Rayleigh) resolution formula
where the retrofocal spot size 
 is proportional to $\lambda$ and the
distance to the TRM,
and inversely proportional 
to the aperture $A$. 
%The Rayleigh criterion is valid only for illumination
%by an incoherent light source \cite{BW}. 
\subsection*{Anomalous Focal Spot-size} 
First we consider the situation where there may be
an inertial range behavior. This requires 
\beq
\label{cond1}
k^2 \gamma^{-2}D_*(\infty) L\gg 1
\eeq
where $D_*(\infty)=\lim_{r\to\infty} D_*(r)$.

In the presence of random inhomogeneities the retrofocal spot
size is determined by ${\cP}_0$ or $T_{\rm tr}$ depending
on which has a smaller support.   For the power-law spectrum we   have the inertial range asymptotic
\beqn
\lefteqn{{T}_{\rm tr}(\bx)}\\
& \sim &\exp{\lt[ - C_{*}^2 \ks^2 L
|\bx-\bx_0|^{2H_*}\gamma^{-2} (4H_*+2)^{-1}\rt]} 
\eeqn
for 
$\ell_0\ll |\bx-\bx_0| \ll L_0.$
Under the following condition
\beq
\label{cond2}
(C_*k\gamma^{-1} \sqrt{L})^{1/H_*} \gg k\gamma^{-1}L^{-1}A\sim \rho_0^{-1}
\eeq
the function $T_{\rm tr}$ is
much more sharply localized around $\bx_0$ than
$\cP_0$.  We define 
the turbulence-induced time-reversal
resolution as
\[
\rho_{\rm tr}=\sqrt{\int |\bx-\bx_0|^2 T^2_{\rm tr}(\bx-\bx_0)d\bx/\int T^2_{\rm tr}(\bx)d\bx}
\]
which 
then has the inertial range asymptotic 
\beq
\label{99}
\rho_{\rm tr}\sim  
\left(\frac{\gamma \lambda} 
{C_{*} \sqrt{L} }\right)^{1/H_*},\quad \ell_0\ll \rho_{\rm tr} \ll L_0.
\eea
 The nonlinear law (\ref{99})
is valid only down to the inner scale $ \ell_0$
below which the linear law prevails 
$
\rho_{\rm tr}\sim \gamma\lambda {L}^{-1/2}.$
We see that under (\ref{cond1})-(\ref{cond2}) $\rho_{\rm tr}$ is independent of 
the aperture, has a
superlinear dependence on the wavelength
in the inertial range. Moreover, the resolution
is enhanced 
as the distance $L$ and random inhomogeneities ($C_*$)  increase. This effect can be explained
by the notion of {\em turbulence-induced aperture}  which
enlarges as $L$ and $C_*$ increase because
the TRM is now  able to capture signals initially
propagating in the more  oblique directions
(see more on this below).
\commentout{
In the limiting case $H_*=1$  the linear
dependence on the wavelength is recovered
\bea
\rho_{\rm tr}^{smooth}  \sim 
\frac{\gamma \lambda} 
{C_{*} \sqrt{L} }
\label{smooth}
\eea
which has a similar form to the  linear law of superresolution 
but is derived under different physical conditions.
 }
 
To recover the linear law previously reported in \cite{BPZ}, let us
consider the situation where $\rho_{\rm tr}=O(\gamma)$
and 
take the limit of vanishing Fresnel number $\gamma\to 0$ in eq. (\ref{9}) by setting $\bx=\gamma \by$. Then we have
\beq
\lim_{\gamma \to 0} D_*(\gamma \by)&=&D_0|\by|^2\\
  D_0&=&\frac{1}{2}\int \Phi(0,\bq)|\bq|^2 d\bq. \nn
 \eeq
 The resulting mean
 retrofocused field $\lan \Psi_{\rm tr}(\by)\ran$ is
 Gaussian in the offset variable $\by$ and the
 refocal spot size on the original scale  is given by
 \[
 \rho_{\rm tr}\sim \gamma \lambda (D_0L)^{-1/2}.
  \]
  Hence the linear law prevails in the sub-inertial range.
\section*{Turbulence-induced aperture and duality}\label{sec:length}
Intuitively speaking, the turbulence-induced aperture
is closely related to how a wave is spread
in the course of  propagation through the turbulent medium.
A
quantitative estimation can be given  by
analyzing
  the spread of wave energy.

Let us calculate the mean energy density in $
{\lan |\Psi(z,\bx)|^2\ran }$
with the Gaussian  initial wave amplitude 
\[
\Psi(0,\bx)=\exp{\lt[-|\bx|^2/(2\alpha^2)\rt]}.
\]
We obtain 
\bea
\nn
&&{\lan |\Psi(L,\bx)|^2\ran }
  =\lt( \frac{\alpha}{2 \sqrt{\pi}}\rt)^d
   \int e^{-|\bw|^2 [\alpha^2/4+\gamma^2 L^2/(4\ks^2
\alpha^2)]}\\
&&\times
     \exp{\lt[-\ks^2/(2\gamma^2) L\int_0^1 {D_*}(\gamma L\bw s/\ks)ds\rt]}
	 e^{i \bw \cdot \bx} \, d\bw.\label{Meq}\nn
\eea
Hence the turbulence-induced broadening can be
identified as convolution with the kernel which is the inverse Fourier transform $\cF^{-1}T$ of the
transfer function 
\[
T(\bw)=   \exp{\lt[-\ks^2/(2\gamma^2) L\int_0^1 {D_*}(\gamma L\bw s/\ks)ds\rt]}.
	 \]
	 In view of (\ref{ttr}), we obtain
	 that 
	 \[
	 \cF^{-1}T(\bx)=\frac{k^2}{\gamma^2 L^2}\cF^{-1}T_{\rm tr}(\frac{k\bx}{\gamma L}).
	 \]
We define the turbulence-induced forward spread $\sigma_*$	
as
\[
\sigma_*=\sqrt{\int |\bx|^2 T^2(\bx) d\bx/\int T^2(\bx)d\bx}
\]
which together with $\rho_{\rm tr}$ then satisfies
the uncertainty inequality:
\beq
\label{dual}
\sigma_* \rho_{\rm tr}\geq \frac{\gamma L}{k}.
\eeq
The equality holds when $T_{\rm tr}$ is Gaussian, i.e.
when $\rho_{\rm tr}\leq \ell_0$  or $\ell_0\ll\rho_{\rm tr}
\ll L_0$ with
$H_*=1$. This strongly suggests the definition of
the turbulence-induced  aperture as
$
A_*=\gamma \, {\lambda L}/ {\rho_{\rm tr}}
$
which is entirely analogous to (\ref{ray}).

Because the coherence  length of the  wave field is closely related to the spread, it is not surprising
then to find that 
the turbulence-induced (de)coherence length $\delta_*$
associated with $\lan \overline{\Psi(L,\bx)}\Psi(L,\by)\ran$
is directly related to $\rho_{\rm tr}$. Indeed, 
one can show that
$\delta_*\approx \rho_{\rm tr}$
when the effect of the turbulent medium 
is dominant over diffraction \cite{tire-siam}. 
\section*{Discussion}
In summary,  we have shown for the parabolic Markovian model 
that, first, the time reversal
resolution can be aperture independent and depend on the wavelength
in a nonlinear way. This is due to the self-similar nature of the media.
Second, we prove an uncertainty inequality for random media where
the conjugate variables are the forward wave spread and the time-reversal
resolution. The equality is attained when the wave structure function $T_{\rm tr}$ is Gaussian.

The preceding analysis has been carried out for a narrow-band
signal. Because of the linearity of the equation  a wide-band signal $u_0(t,\bx)$  can be decomposed into frequency components
each of which can be analyzed as above and
then resynthesized. The mean retrofocused signal can 
be calculated as
\bea
\lefteqn{  \lan u_{\rm tr}\ran (\tau,\bx) }\\   &=&\frac{1}{2\pi\gamma^2 L^2}\int\overline{u_0(t,\by)}
\int \hat{\II}_A\lt(\frac{k(\bx+\by)}{\gamma L}\rt)e^{-i k(t+\tau)} \nn\\
&&\times e^{i k|\bx|^2/(2\gamma L)} e^{-i k|\by|^2/(2\gamma L)}k^2 T_{\rm tr} (\bx-\by)dk dt \nn
\eea
from which it follows that the turbulence-induced
spread in time is given by convolution with a 
{\em Gaussian } kernel because $T_{\rm tr}$ is
Gaussian in $k$, see (\ref{ttr}). The Gaussian kernel has
a offset-$\bx$-depending variance 
$
\sigma^2_{\rm tr}(\bx)={L}
\int^1_0 D_*(s \bx)ds/{\gamma^2}
$
which grows rapidly with the offset if $L\gg 1, \gamma\ll 1$.
It is precisely this rapid change of temporal dispersion rate
with the offset that produces the sharp spatial retrofocusing
of the time-reversed pulse.

Our results above have been limited to the mean value
of the time-reversed retrofocused field. 
 Its second or
higher moments can be determined from those of
the Wigner distribution which are not exactly solvable.
However, the mean field is sufficient for determining all
the higher moments in case of self-averaging. 
Self-averaging occurs, for example, when the narrow-band beam
width in the transverse directions is large compared to the correlation length of
the random medium or when the signal is wide-band
\cite{Fink}. The former case has been analyzed
extensively in the literature (see \cite{rad-arma}
and references therein) and there arise several canonical
radiative transfer equations as the self-averaging scaling limits.
The latter case of wide-band
signals has only
been studied for the one-dimensional medium where
the issue of spatial focusing does not arise,  see
\cite{fouque}. 
In the  near-self-averaging regime the second
moment of the Wigner distribution can be calculated
and will be reported elsewhere.

{\bf Acknowledgment}. 
The research supported in part by ONR Grant N00014-02-1-0090,
DARPA Grant N00014-02-1-0603. 
A. F. is supported in part  by
NSF grant DMS 0306659 and
 K. S. 
by NSF grant DMS0307011 and the Sloan Foundation.

%\bibliography{tire-prl}

%\end{document}

\end{document}